Article

# Decarbonization without deindustrialization in future pathways for European industry

## Graphical abstract

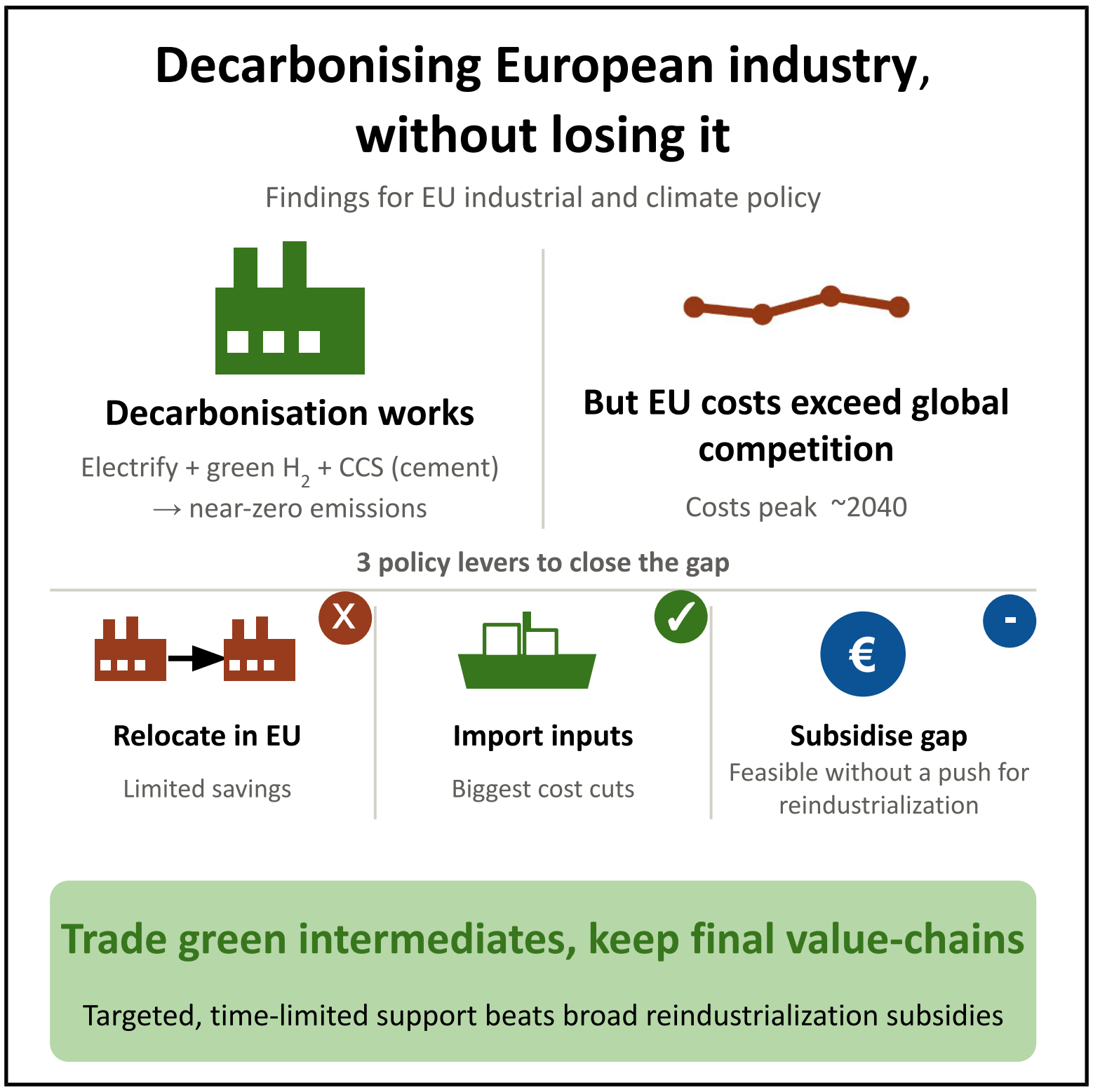

## Highlights

- Deep industrial decarbonization in Europe is feasible, driven by electrification
- Cost competitiveness peaks around 2040 during capital-intensive investment phases
- Importing green intermediates cuts system costs more than domestic relocation
- Large-scale reindustrialization would require fiscally unsustainable subsidies

## Authors

Alice Di Bella, Toni Seibold, Tom Brown, Massimo Tavoni

## Correspondence

alice.dibella@cmcc.it

## In brief

Using an extended PyPSA-Eur energy system model, this study examines whether European industry can decarbonize without losing competitiveness. Results show that while electrification and green hydrogen make deep decarbonization feasible, importing green intermediate goods offers a more cost-effective path than relocation or large subsidies.

Article

# Decarbonization without deindustrialization in future pathways for European industry

Alice Di Bella,[1,2,3,5,*] Toni Seibold,[4] Tom Brown,[4] and Massimo Tavoni[1,2,3]
[1]Department of Electronics, Information and Bioengineering, Politecnico di Milano, Milano 20133, Italy
[2]CMCC Foundation - Euro-Mediterranean Center on Climate Change, Milan 20144, Italy
[3]RFF-CMCC European Institute on Economics and the Environment, Milan 20144, Italy
[4]Department of Digital Transformation in Energy Systems, Institute of Energy Technology, Technische Universität Berlin, Berlin 10623, Germany
[5]Lead contact
*Correspondence: alice.dibella@cmcc.it

**SCIENCE FOR SOCIETY** Europe has pledged to reach climate neutrality by 2050, but heavy industry (e.g., steel, cement, and chemicals) remains challenging to clean up. Policymakers worry that the costs of going green could push these industries out of Europe, undermining jobs and economic security just as climate goals are being met. This study uses a detailed model of Europe's energy and industrial systems to examine whether decarbonization and competitiveness can be reconciled and to identify the effective pathways to do so. The findings show that emissions reductions in heavy industry are feasible, mainly through electrification and clean hydrogen. However, the transition includes a sharp cost increase around 2040 as new low-carbon plants and equipment are built. During this period, production in Europe becomes more expensive than producing in regions with lower-cost renewable energy. Three policy responses are compared: relocating factories to sunnier or windier parts of Europe, importing partially processed low-carbon materials from abroad while keeping finished-goods production at home and subsidizing domestic producers directly. Relocating factories within Europe yields relatively limited benefits; wide-ranging subsidies to rapidly expand low-carbon industrial capacity would require public spending far beyond current European programs, raising concerns about affordability and fairness. Importing green intermediate materials, by contrast, offers a more effective middle path: it lowers costs substantially while enabling Europe to retain higher-value manufacturing steps and the associated jobs. Overall, the results suggest that European industrial policy is likely to be more successful, and fiscally sustainable, if it focuses on protecting and upgrading existing industrial capacity through targeted, time-limited support and strategic trade partnerships rather than pursuing a subsidy-heavy expansion of energy-intensive production at home.

## SUMMARY

Decarbonizing European industry while maintaining international competitiveness are two central objectives of European Union (EU) policy, yet they may conflict. This study investigates trade-offs between these goals across alternative industrial production pathways, from continued decline to policy-driven reindustrialization, focusing on steel, cement, methanol, ammonia, and high-value chemicals. Using an extended PyPSA-Eur energy system model with detailed process-level representation, we compare domestic low-carbon production costs against imported alternatives and evaluate three policy strategies: intra-European relocation, selective green intermediate imports, and targeted subsidies. Deep industrial decarbonization is technically feasible, with electrification playing a central role, but cost competitiveness peaks during the investment-intensive transition phase around 2040. Intra-European relocation delivers limited savings and faces significant real-world constraints. Strategic imports of green intermediates substantially reduce system costs while preserving downstream industrial activity. Overall, European industrial policy should prioritize decarbonizing existing capacity, leveraging green intermediate imports to retain high-value domestic production, supported by targeted, time-limited public support.

## INTRODUCTION

The European Union (EU) has committed to achieving climate neutrality by 2050, requiring rapid and substantial reductions in greenhouse gas (GHG) emissions across all economic sectors.[1] Industry plays a central role in this transition, as many existing industrial processes remain incompatible with deep decarbonization pathways. In response, the EU has introduced a range of policy initiatives, including the Net-Zero Industry Act and the Clean Industrial Deal, aimed at accelerating clean technology deployment while strengthening industrial resilience and strategic autonomy.[2,3] At the same time, concerns over high energy prices, regulatory complexity, and slower innovation have raised questions about Europe's ability to maintain industrial competitiveness during the transition.[4] European industrial products are already facing structural cost disadvantages in global markets, with energy-intensive goods often priced above international competitors, reinforcing fears of deindustrialization and carbon leakage. Recent policy discourse increasingly frames decarbonization not only as a climate imperative but also as a potential driver of long-term competitiveness and economic security.[3]

In 2021, the EU27 industrial sector emitted approximately 600 Mt of $CO_2$, accounting for around 20% of total EU GHG emissions.[5,6] Emissions are concentrated in a small number of energy- and material-intensive sectors, notably iron and steel, chemicals, and non-metallic minerals, which together account for more than two-thirds of industrial emissions. These hard-to-abate sectors face structural challenges due to their reliance on fossil fuels for high-temperature heat and carbon feedstocks, as well as the limited maturity of low-carbon alternatives. Electrification, green hydrogen, and carbon capture and storage (CCS) are widely recognized as key technological pillars of industrial decarbonization. Recent reviews emphasize that industrial decarbonization requires coordinated technological portfolios, long investment cycles, and supportive policy frameworks to address both technical and economic barriers.[7–11]

A growing body of literature has employed energy system models (ESMs) and integrated assessment models (IAMs) to explore pathways for industrial decarbonization, assessing technology deployment, sectoral interactions, and emissions reductions.[12–16] For example, Pehl et al.[16] extend the REMIND integrated assessment model with a detailed representation of industrial energy demand and technology pathways, enabling the analysis of industrial decarbonization within a global energy-economy-climate system context. Other studies focus on sector-specific modeling of industrial transformation. Neuwirth et al.[15] develop a site-specific model of hydrogen-based steel and chemical production in Europe, while Raillard-Cazanove et al.[13] investigate cost-optimized decarbonization pathways for key industrial sectors using detailed process representations. Together, these studies highlight the technical feasibility of electrification, hydrogen use, and carbon capture technologies, while also emphasizing uncertainties related to costs, data availability, and technology performance.[17–19]

Another strand of literature approaches industrial decarbonization from an economic and competitiveness perspective, focusing on its implications for industrial costs, market structures, and supply chains. Rinaldi et al.[20] employ a hybrid input-output modeling framework to assess the environmental and market implications of alternative steel decarbonization strategies in the EU. Similarly, Di Foggia and Beccarello[21] analyze decarbonization strategies in the European steel sector with a focus on investment risks, policy incentives, and industrial competitiveness. Recent policy-oriented contributions emphasize that industrial decarbonization entails significant economic and distributional risks, highlighting the need for pathways that align emissions reductions with competitiveness, and thus for policy-relevant assessments that go beyond technology cost optimization.[11,22]

Further research examines industrial decarbonization in a global context and highlights the role of geographical differences in renewable energy resources. Verpoort et al.[23] analyze how heterogeneous renewable resource availability may reshape global production patterns for energy-intensive commodities such as steel, urea, and ethylene, potentially leading to new green industrial value chains. Similarly, Trollip et al.[24] explore how renewable-rich regions such as South Africa could become competitive exporters of green primary iron, while Wang et al.[25] investigate the potential for renewable hydrogen and iron ore resources to support green steel production in Australia. In parallel, several analyses suggest that Europe could reduce system-wide decarbonization costs by increasing imports of energy carriers or energy-intensive goods from regions with abundant renewable resources, including green hydrogen, synthetic fuels, electricity, and steel.[24,26,27]

Neumann et al.[28] combine a global energy supply-chain model with the PyPSA-Eur framework to show that importing renewable energy and steel can reduce Europe's infrastructure requirements and overall system costs. While such strategies may improve cost efficiency, they also raise concerns regarding deindustrialization, employment losses, innovation capacity, and increased dependence on external suppliers. These trade-offs are particularly relevant in the current policy context, where industrial policy, energy security, and climate objectives are increasingly intertwined.

Despite extensive modeling work on industrial decarbonization and growing attention to policy risks and competitiveness concerns, an important gap remains. Existing studies tend to emphasize either system-level cost optimization or sector-specific technology pathways, while paying limited attention to how decarbonization strategies interact with industrial competitiveness and policy instruments such as subsidies. This paper advances the literature by providing an integrated assessment that combines a detailed, process-level representation of key industrial sectors with economy-wide energy system dynamics, enabling an explicit evaluation of competitiveness trade-offs and policy-relevant transition strategies over the full pathway to net zero.

This study addresses this gap by extending the PyPSA-Eur framework to analyze European industrial decarbonization through a competitiveness-oriented lens. We include a plant-level representation of key industrial sectors, accounting for the type, location, and age of existing facilities, and assess alternative transition strategies within a myopic optimization framework covering the full transition period. Beyond technology deployment, the analysis explicitly considers three policy-relevant mechanisms: (1) intra-European relocation of industrial activity to optimize production and infrastructure use; (2) selective imports of green intermediate goods, such as hot briquetted iron (HBI), methanol, and ammonia; and (3) targeted government subsidies aimed at closing cost gaps with regions benefiting from lower renewable energy costs.

**Table 1. Link between research questions, analyzed mechanisms, and results presented in the paper**

| Research ques-tion | Analyzed mechanisms | Results section | Key figures |
|---|---|---|---|
| RQ1 | technology portfolios, production costs, and competitiveness of domestic green industrial goods | decarbonization and competitiveness | Figures 1, 2, and S1 |
| RQ2 | intra-European relocation of industrial activity and imports of green intermediate goods | industrial relocation and green intermediates trade | Figures 3 and 4 |
| RQ3 | targeted government subsidies and competitiveness gaps relative to green imports | government subsidies | Figure 5 |

By analyzing how these strategies affect production costs, emissions, and trade-offs between domestic production and imports, the paper provides insights into how deep decarbonization can be reconciled with industrial competitiveness under realistic policy constraints. In doing so, it complements existing system-level analyses by shifting the focus toward industrial structure, policy design, and transition feasibility. Throughout, competitiveness is operationalized as cost competitiveness, proxied by model-derived cost gaps, marginal prices, and implied subsidy magnitudes, rather than in the full macroeconomic sense of global market share.

The paper addresses the following research questions:

**RQ1**: how can European industries reduce emissions while maintaining cost competitiveness, and what are the implied cost gaps relative to global production alternatives?
**RQ2**: what role do intra-European relocation and imports of green intermediate goods play in shaping less disruptive, lower-cost transition pathways to net zero?
**RQ3**: to what extent can targeted subsidies close the cost gap and support the retention of green industrial production in Europe?

The remainder of the paper is structured as follows: methods describes the modeling framework and key assumptions. Results presents the results of the scenario analysis. Finally, discussion explores the implications for industrial and climate policy.

## RESULTS

To guide the interpretation of the results, Table 1 provides an overview of how the model outputs presented in this section relate to the research questions and policy mechanisms analyzed in the study. We first assess technology choices and cost competitiveness under the scenarios with climate policy and different industrial production levels (RQ1), evaluate relocation and intermediate imports as alternative strategies (RQ2), and finally quantify the implied subsidy requirements to retain production within Europe (RQ3).

### Decarbonization and competitiveness

#### *Technology portfolios*

Figure 1 presents the cost-optimal technological pathways identified by PyPSA-Eur across scenarios. The first column shows the "no climate policy" scenario under stabilization of industrial production, while the remaining columns depict "climate policy" scenarios under continued decline, stabilization, and reindustrialization. Methanol production increases in all the climate policy scenarios to meet growing maritime fuel demand. All optimizations assume no relocation within Europe and no intermediate goods imports. For hydrogen produced via electrolysis, the distinction between "green" and "gray" electricity is derived from the overall electricity mix in the system, reflecting the composition of total power generation rather than individual production processes.

Across scenarios, industrial decarbonization is primarily driven by electrification and green hydrogen deployment (Figure S1), followed by CCS, particularly for cement. For steel, ammonia, and methanol, green hydrogen-based production becomes dominant from 2040 onward under stringent decarbonization targets (−90% GHG). Steelmaking relies heavily on scrap-based electric arc furnaces (EAF) due to their cost and emissions advantages, reaching the imposed cap of 75%, which reflects quality constraints that still require some primary steel. Steam methane reforming (SMR) with CCS has a small role in 2030, as captured $CO_2$ can be utilized in other industrial processes. In the *reindustrialization* scenario, plastics increasingly rely on sequestered $CO_2$ for Fischer-Tropsch synthesis, reaching approximately 25% of output by 2050, while biomass use remains constrained by competing demands. For cement decarbonization, point-source CCS emerges as the primary mitigation option, reaching full deployment by 2040. Since the only sector-specific abatement option available is CCS retrofit, the model's alternative would be to offset cement emissions through system-wide carbon dioxide removal (CDR) technologies such as direct air capture (DAC) or bioenergy with CCS (BECCS). However, the model consistently favors direct capture at the source, reflecting the lower cost of point-source CCS relative to CDR under the modeled assumptions.

#### *Prices of industrial goods*

Figure 2 reports production-weighted average European industrial commodity prices for *climate policy* scenarios at various

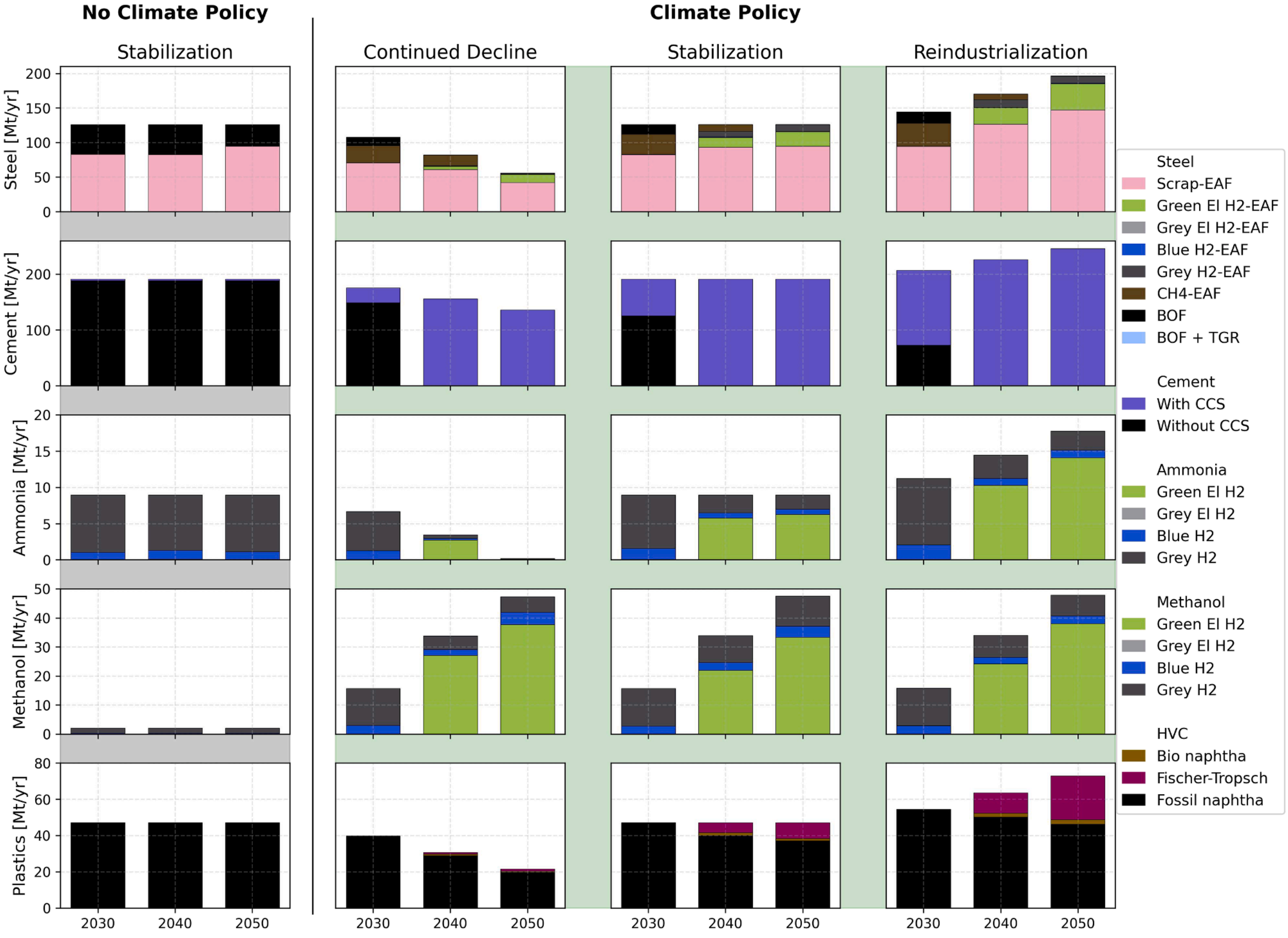

**Figure 1. Production of industrial goods in the four scenarios**

The first column represents the scenario with no climate policy and stabilization of industrial production; the other three columns show scenarios implementing the EU climate policies, spanning from continued decline to stabilization and then reindustrialization. Rows indicate the different industrial sectors and the technologies available in the model are depicted in the legend.

production levels. Prices are derived from Karush-Kuhn-Tucker multipliers, representing marginal production costs by region and time. Historical price ranges and recent European prices are included for reference.

Decarbonization leads to higher industrial production costs, with a pronounced peak around 2040, followed by a decline by 2050 as major investment phases are completed. Cost increases are driven primarily by capital-intensive investments in low-carbon technologies, particularly electrolyzes and higher electricity consumption. For steel, ammonia, and methanol, green hydrogen becomes the dominant cost driver. For cement and plastics, carbon pricing and carbon abatement costs play a larger role.

Comparisons with literature-based cost ranges for imported low-carbon goods from renewable-rich regions indicate that plastics remain competitive across scenarios, while steel competitiveness is maintained only under *continued decline*. Ammonia and methanol remain competitive under both *continued decline* and *stabilization*, reflecting lower marginal cost pressures at reduced production volumes. Overall competitiveness outcomes depend on domestic production trajectories and uncertainties surrounding green import costs. In Figure 2, for methanol and plastics we show the dotted lines when we do not include end-of-life (EOL) emissions, referring to the release of carbon embedded in the product at the end of its use phase. The inclusion or exclusion of EOL emissions depends on the carbon pricing specification, as carbon prices may be applied either only to production emissions or to the full life cycle of carbon embedded in products.

## Industrial relocation and green intermediates trade

To moderate cost increases under deep decarbonization while limiting carbon leakage, we evaluate two strategies: (1) relocation of industrial activity within Europe toward regions with favorable renewable potentials and (2) selective imports of green intermediate goods from outside Europe.

### *Relocation within Europe*

Allowing full relocation of industrial activity within Europe leads to a pronounced concentration of production in regions with abundant and low-cost renewable resources, notably Spain, the

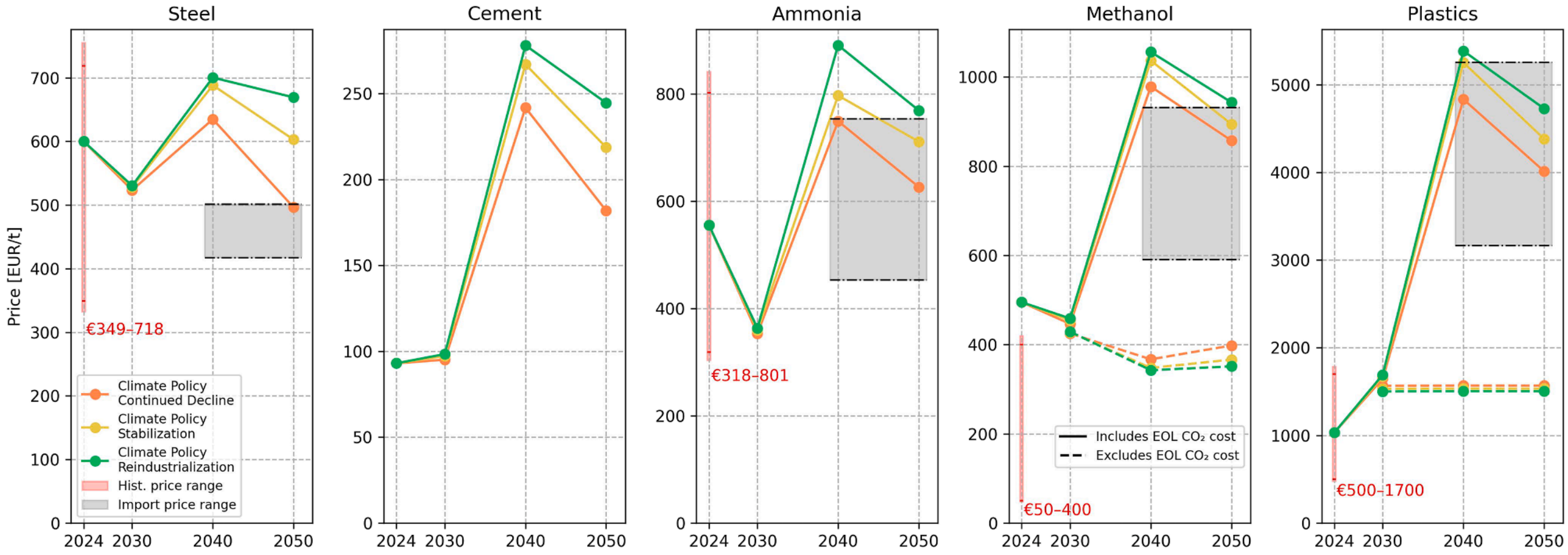

**Figure 2. Prices of industrial goods, average across European countries and time steps, for different scenarios**

Dotted lines for methanol and plastics represent a price when no carbon price on end-of-life (EOL) emissions is applied. The red line indicates historical price ranges, while the gray bar represents the costs for green products from outside the EU's borders.

Nordic countries, and the United Kingdom (Figure 3). This spatial reallocation reflects a strong "renewables pull" effect[29] and results in lower industrial electricity expenditures across all scenarios, as shown in Figure 3C.

However, these results represent a purely cost-optimal outcome that abstracts from several real-world constraints. The model does not account for relocation costs, labor market frictions, regional policy considerations, supply-chain dependencies, or the infrastructure investments required to support large-scale industrial shifts. Quantifying these constraints consistently across sectors and countries remains challenging due to the limited availability of reliable data. For this reason, relocation in the model should be interpreted as a stylized scenario rather than a prediction of future industrial geography. Instead, the exercise provides a benchmark of the potential energy-system cost savings associated with relocating production to regions with more favorable renewable resources.

### *Imports of green intermediate goods*

As an alternative to large-scale relocation, we assess the impact of importing green intermediate goods, HBI, ammonia, and methanol, from outside Europe starting in 2040. This strategy aims to reduce exposure to high domestic production costs while preserving downstream industrial activity within Europe. Import prices are taken from literature estimates[28] that evaluate low-carbon export costs in renewable-rich regions after satisfying domestic energy demand. Figure 4 presents the resulting technology shares and system-level cost outcomes for the *continued decline* and *reindustrialization* scenarios.

Imports of green intermediates substantially reshape the European industrial technology mix from 2040 onward. Steel production increasingly combines imported HBI with domestic EAF, while ammonia and methanol production shift almost entirely toward imports by 2050. This transition reduces domestic green hydrogen demand by approximately 25% and leads to a significant reduction in annual system costs and carbon prices, particularly under the *reindustrialization* scenario.

Compared with intra-European relocation, the import of green intermediates achieves larger cost reductions while limiting structural disruption to existing industrial capacity. By retaining final-stage processing within Europe while outsourcing the most energy-intensive production steps, this strategy can reduce exposure to high domestic energy costs while preserving downstream industrial activity. These results suggest that imports of green intermediates may act as a cost-containment mechanism during the transition, particularly in periods when European renewable capacity remains constrained.

### Government subsidies

Figure 5 shows the average annual subsidies required to retain industrial production within Europe between 2035 and 2055. Subsidy needs vary significantly across sectors and scenarios.

The highest subsidy requirements arise under *reindustrialization* without intermediate imports, reaching approximately 235 bnEUR/year. Plastics exhibit high total subsidy needs primarily due to large production volumes rather than lack of price competitiveness. Ammonia, methanol, and steel require more moderate support, particularly when intermediate imports are allowed.

## DISCUSSION

This study has the goal to examine whether deep industrial decarbonization in Europe can be achieved while preserving cost competitiveness and industrial capacity. This tension is increasingly central in EU climate and industrial policy discussions, where net-zero targets are pursued alongside competitiveness and resilience objectives.[3,30] Beyond identifying this tension, the analysis explicitly evaluates which combinations of industrial policy instruments, subsidies, trade in intermediates, and spatial reallocation, remain fiscally, legally, and politically feasible under EU institutional constraints.

At high level, the results confirm a central tension already highlighted in the literature: although low-carbon strategies,

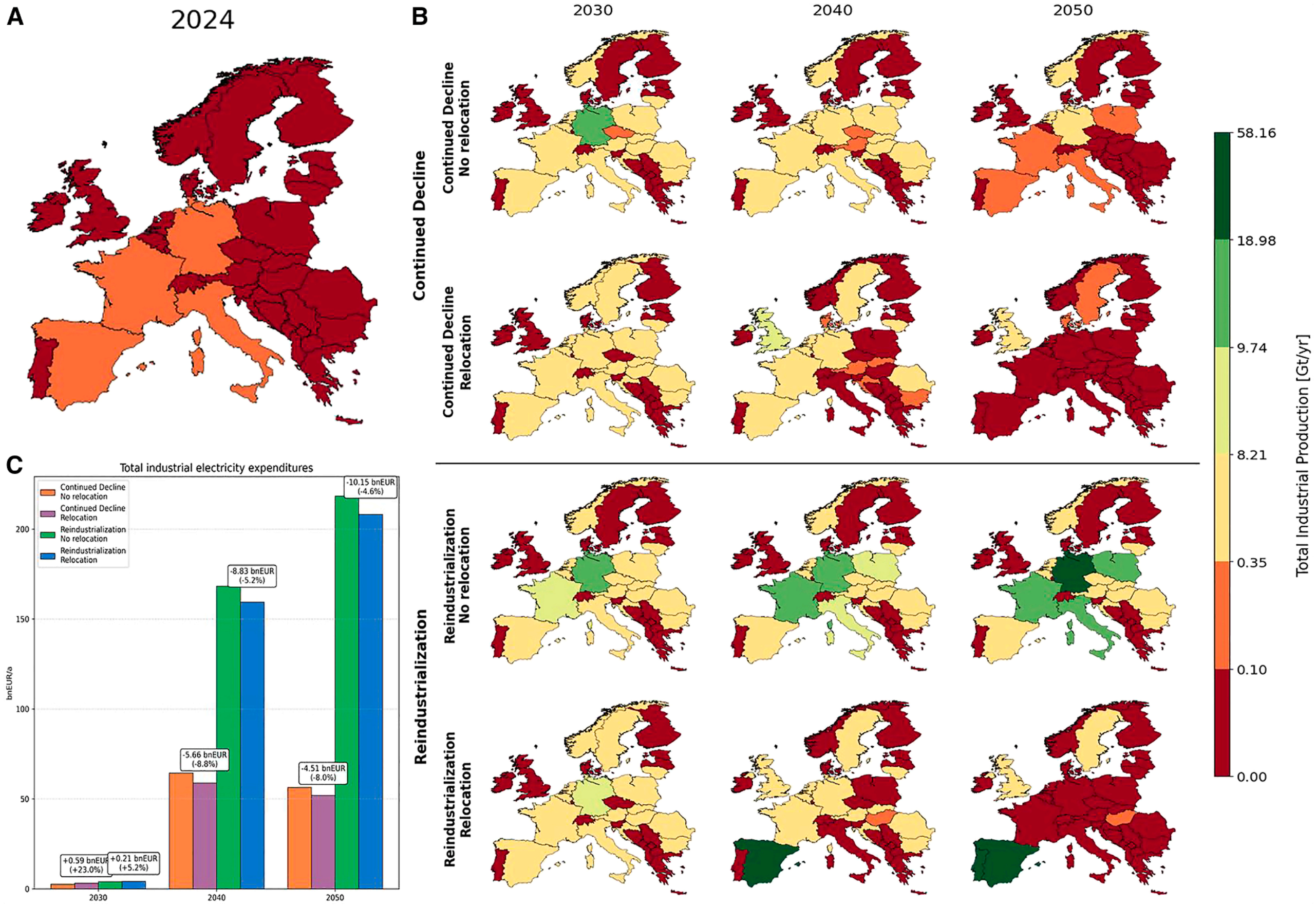

**Figure 3. Impacts of relocation within Europe of industrial hubs**

(A) Total industrial production levels across European countries in 2024, showing current capacity distribution.

(B) Projected industrial production trajectories for 2030, 2040, and 2050 under four scenarios: *continued decline* (two top rows) with *no relocation within Europe* and with *relocation within Europe*, and *reindustrialization* (two bottom rows), again with and without relocation. Color scale represents total industrial production in Gt/a.

(C) Industry expenditures for electricity across scenarios and years, in billion euros per year. Boxes report absolute and percentage differences relative to the *no relocation* case.

particularly electrification and green hydrogen use, enable substantial emissions reductions, their large-scale deployment increases production costs relative to regions with abundant and inexpensive renewable resources.

The analysis provides additional insights on cost dynamics. Cost pressures peak around 2040, when capital-intensive investments in low-carbon process technologies coincide with tightening emissions constraints, but decline by 2050 as capital stock turnover is completed, although price differentials with renewable-rich regions persist. This finding is consistent with previous system-level studies such as Victoria et al.[31] and Nijsse et al.,[32] as well as sectoral modeling by Raillard-Cazanove et al.,[13] which highlight the importance of investment cycles and transition dynamics. At the same time, our results reinforce findings from industrial decarbonization studies such as Madeddu et al.[12] and Raillard-Cazanove et al.,[13] which identify electrification, green hydrogen, and point-source CCS as central drivers of emissions reductions, while showing that the associated cost increases are concentrated in specific phases of the transition rather than uniformly distributed over time.

From an industrial policy perspective, the results suggest that strategies centered on large-scale reindustrialization under climate targets would require substantial and sustained public support. In the most ambitious scenarios, annual subsidy requirements exceed 200 bnEUR, far surpassing historical EU renewable energy subsidies (around 80 bnEUR/year in 2021[33]) and current funding ambitions under initiatives such as the Clean Industrial Deal (approximately 100 bnEUR[3]). Such magnitudes raise concerns regarding fiscal feasibility and compatibility with EU competition law and state aid rules. State aid that distorts competition and affects trade within the internal market is in principle prohibited and may only be authorized under strict conditions. While recent frameworks allow support for decarbonization and clean technologies,[34] public support must be necessary, proportionate, and time limited, and cannot permanently offset structural cost disadvantages. The subsidy levels needed for a major reindustrialization would therefore face legal

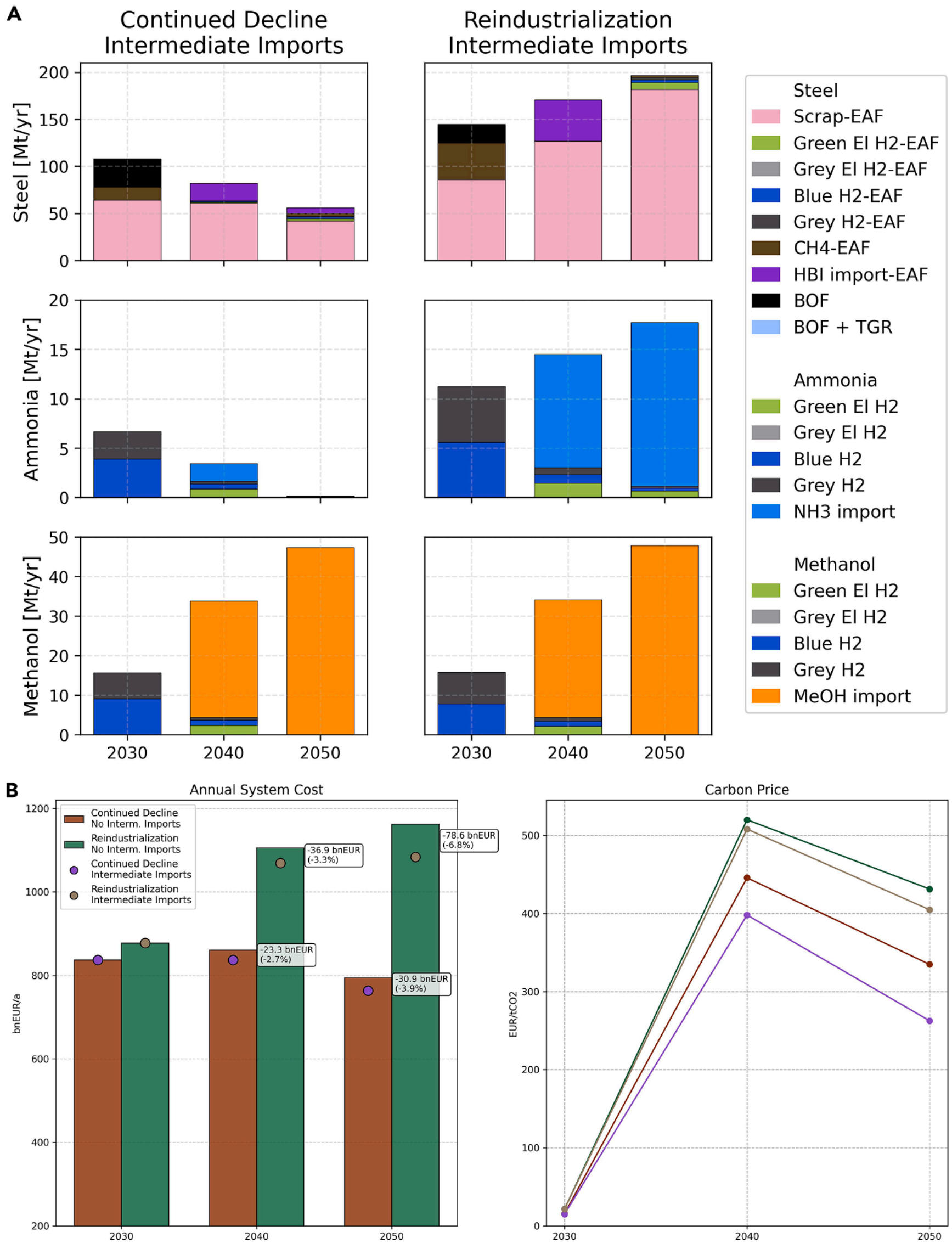

**Figure 4. Impacts of green intermediate industrial imports**

(A) Projected industrial production trajectories for 2030, 2040, and 2050 under two scenarios with intermediate imports: continued decline and reindustrialization.

(B) Annual system costs (bnEUR/a) and carbon prices (EUR/t$CO_2$), compared with the corresponding no intermediate imports cases. Boxes report absolute and percentage differences relative to the no intermediate imports scenario.

and institutional barriers, particularly if concentrated in a small number of energy-intensive sectors or member states, and risk exacerbating fragmentation of the internal market. These findings imply that subsidies can only play a transitional and selective role, rather than forming the backbone of a long-term European reindustrialization strategy under net-zero constraints. Importantly, this analysis extends the existing literature by explicitly quantifying the scale of subsidies required to maintain industrial competitiveness.

By contrast, more moderate industrial strategies emerge as comparatively robust and policy relevant. Allowing selective imports of green intermediate goods, such as HBI, ammonia, and methanol, reduces system costs while preserving downstream industrial activity within Europe. This approach enables Europe to retain value-added stages of production, limit exposure to volatile energy prices, and reduce the scale of public subsidies required to maintain competitiveness. These findings are consistent with previous work by Fasihi et al.,[26] Trollip et al.,[24] and Neumann et al.,[28] which highlight the potential for cost reductions through international trade in energy carriers and energy-intensive goods. From a policy perspective, this suggests a shift from protecting entire value chains toward selectively safeguarding high-value-added stages, while allowing energy-intensive upstream processes to partially internationalize under controlled trade and carbon pricing regimes.

Intra-European relocation of industrial activity can reduce production costs by shifting energy-intensive processes toward regions with abundant and lower-cost renewable electricity, reflecting a clear “renewables pull.” However, the results also confirm that relocation alone offers limited savings relative to the scale of the transition. While energy expenditures decline, the model abstracts from broader social, economic, and political frictions associated with relocation, including employment effects, regional disparities, supply-chain reconfiguration, and infrastructure constraints. Consistent with recent assessments, including Verpoort et al.,[23] these findings suggest that geographical optimization alone may not fully capture the constraints shaping industrial transformation. As a result, relocation within Europe is unlikely to constitute a dominant decarbonization strategy on its own, and trade in intermediate goods may provide a more effective mechanism to balance decarbonization and competitiveness.

Taken together, the results point toward a pragmatic policy direction in which European industrial policy prioritizes the decarbonization and retention of existing capacity rather than aggressive expansion. Strategic use of intermediate imports, combined with targeted, time-limited subsidies, appears more compatible with fiscal and institutional constraints than broad-based support for full domestic production of all energy-intensive intermediates. In this context, revenues from the EU Emissions Trading System and the Carbon Border Adjustment Mechanism could help finance transitional support, provided distributional impacts are carefully managed through instruments such as the Social Climate Fund.

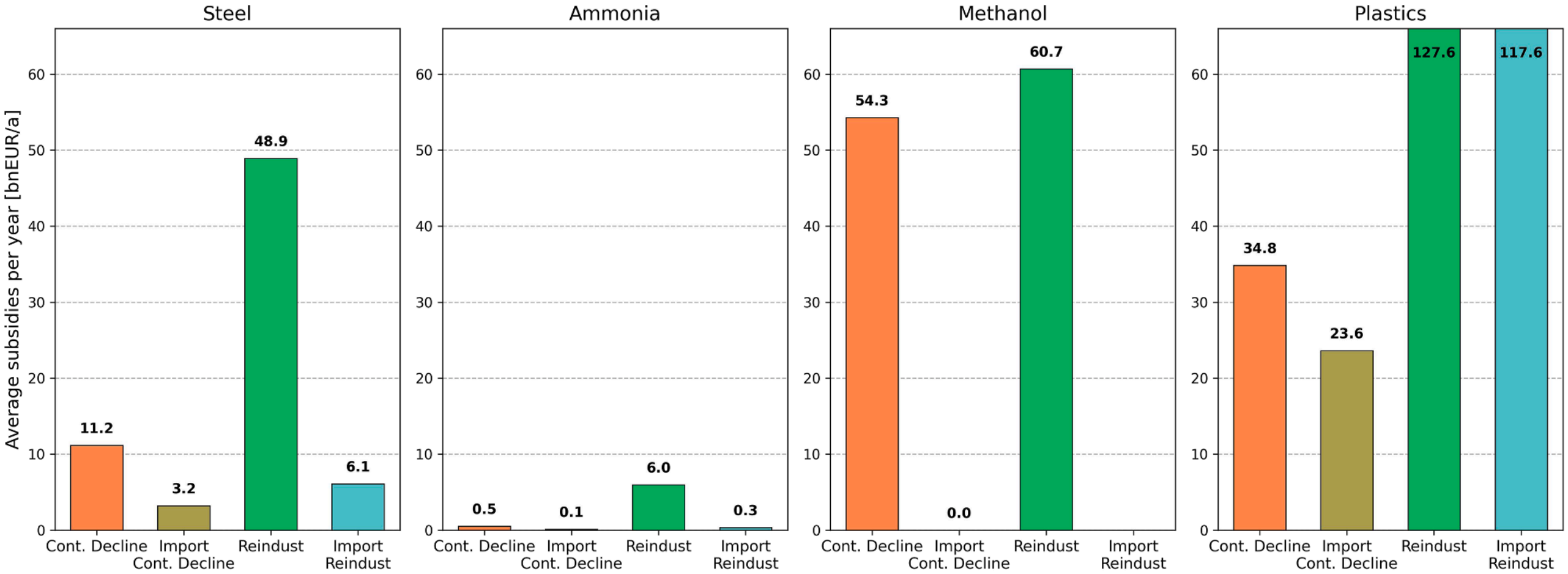

**Figure 5. Average annual subsidy requirements (bnEUR/a) for the five industrial sectors under four policy scenarios: *continued decline*, with and without *intermediate imports*, and *reindustrialization* with and without *no intermediate imports*.**

More broadly, the analysis underscores that industrial decarbonization cannot be assessed solely through a technological or cost-optimization lens. Trade-offs between emissions reductions, competitiveness, fiscal capacity, and political feasibility are intrinsic to the transition. By explicitly modeling these interactions over the full pathway to net zero, this study contributes to a more policy-relevant understanding of how Europe might navigate the industrial transformation ahead.

### Limitations and future research directions

This study has some limitations that open opportunities for future research. We do not consider the possibility to replace today's plastics primary production with higher recycling rates, and we omit biomass-based methanol to maintain tractable $CO_2$ accounting and for the sake of simplicity. Including these pathways in future analyses would expand decarbonization options and enable a more comprehensive assessment of interactions with other low-carbon strategies. This study's technology portfolio is not fully comprehensive, omitting CCS at $CH_4$-EAF steel plants, clinker-to-output improvements in cement, and other measures requiring more detailed modeling. It also excludes $CO_2$ transport and storage infrastructure, which could affect CCS and CDR adoption. Despite these gaps, major low-carbon options are captured, and future work incorporating sensitivities and $CO_2$ infrastructure could refine assessments of industrial decarbonization pathways. The analysis also omits potential geopolitical shocks and their effects on commodity and technology costs, particularly for clean energy technologies dependent on concentrated critical raw materials. Future work incorporating cost sensitivities and supply-chain risks would help evaluate the robustness of decarbonization pathways under market and geopolitical uncertainties.

## METHODS

### Model development

The ESM employed in this study is PyPSA-Eur, an open-source, high-resolution modeling tool developed to explore cost-optimal decarbonization pathways for the European energy system. For a comprehensive description, readers are referred to the official model documentation,[35] and the GitHub repository.[36] PyPSA-Eur is widely used[37–39] and enables integrated analyses of electricity, heating, transport, industry, agriculture, shipping, and aviation. Figure 6 offers a scheme of the sectors and links between them. It was chosen for its detailed power grid representation and ability to capture inter-sector interactions. The model performs linear optimizations to minimize system costs, optimizing investments and operations across generation, storage, conversion, and transmission. Investment and dispatch decisions are optimized jointly across electricity generation, storage, and end-use sectors including industry, heating, and transport, allowing for endogenous trade-offs between sectors. This structure makes PyPSA-Eur particularly suitable for analyzing interactions between industrial decarbonization pathways, energy infrastructure, and policy constraints.

Industrial modeling in PyPSA-Eur, detailed in Victoria et al.,[37] uses Joint Research Centre's Integrated Database of the European Energy System (JRC-IDEES) data[6] to estimate energy demands and process emissions per unit output, with exogenous assumptions on low-carbon technology uptake. Production volumes are held constant until 2050, yielding predetermined electricity, hydrogen, biomass, and oil demands. Neumann et al.[28] extended the model to include material imports and EAF for steel, but their greenfield 2050 optimization ignores the spatial distribution and operational status of existing plants.

While the current framework of PyPSA-Eur is able to capture industrial energy demand, this study requires a more detailed representation of sectors and technologies. To this end, PyPSA-Eur was extended to represent in detail five key sectors, iron and steel, cement, ammonia, methanol, and chemicals. By modeling specific processes and techno-economic parameters, the model endogenously determines least-cost technology pathways and allocates mitigation across sectors to achieve economy-wide net-zero emissions. The model development includes the representation of industrial processes, with

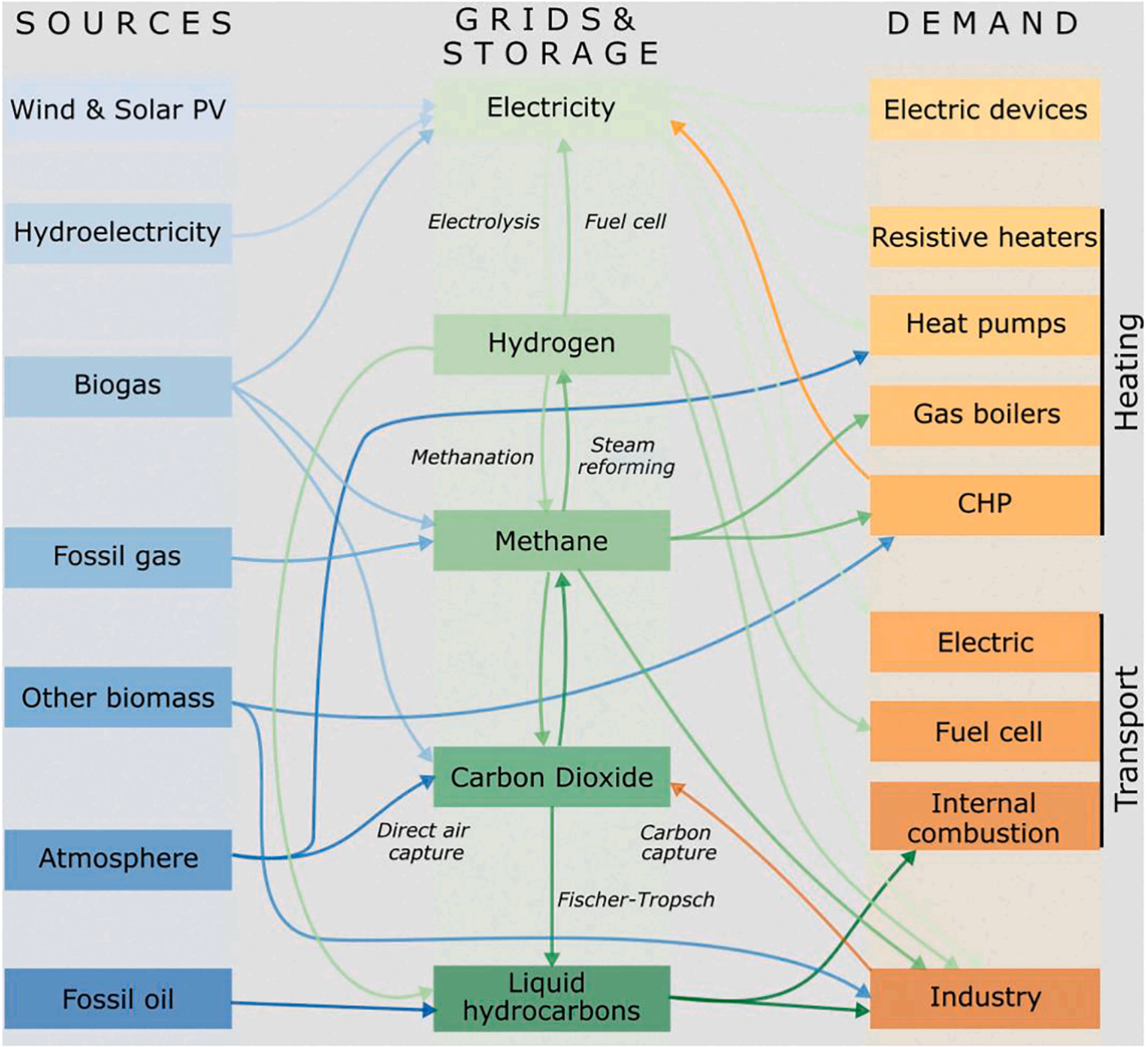

**Figure 6. Diagram PyPSA-Eur sectors and the links between them**

The scheme is present in PyPSA-Eur documentation page.[35]

technology-specific energy inputs, emission factors, capital costs, and operational constraints. Existing industrial facilities are spatially allocated using plant-level datasets,[40,41] which provide information on plant location, technology type, and commissioning year. Data on existing capacity are aggregated at the model node level (in this study corresponding to national nodes). Plant age is represented as a capacity-weighted average of the commissioning years of plants within each node and technology. Decommissioning of existing capacity is then determined based on the estimated lifetime of each technology. High-temperature industrial technologies are modeled with minimum partial-load constraints reflecting their limited operational flexibility.

In the iron and steel sector, the model represents three production routes: the conventional blast furnace-basic oxygen furnace (BOF), which uses iron ore, coking coal, and electricity; natural gas-based direct reduced iron coupled with an electric arc furnace ($CH_4$-EAF); and hydrogen-based direct reduction coupled with an EAF ($H_2$-EAF). Additionally, steel scrap can be used as a substitute for primary iron in EAF routes, with scrap availability constrained by scenario-specific projections. BF-BOF can be decarbonized through top gas recycling (TGR) with post-combustion CCS. In the cement sector, the model represents the conventional dry-process rotary kiln, fired by natural gas, as the only production route. The sole decarbonization option for existing plants is a post-combustion CCS retrofit. For ammonia, production occurs via the Haber-Bosch process, which synthesizes ammonia from hydrogen and nitrogen; hydrogen can be produced from the power system (electrolysis) or from SMR, with or without CCS. For methanol, the primary production route is catalytic metabolization, combining hydrogen with $CO_2$ captured from the atmosphere or industrial point sources. For high-value chemicals (HVC, representing plastics feedstocks), the primary production route is the conventional naphtha steam cracker, which converts oil into HVC with associated electricity inputs and EOL $CO_2$ emissions from product degradation. The oil feedstock can be of fossil origin, synthetically produced via the Fischer-Tropsch process (which yields a slate of hydrocarbon products, with naphtha used here as a proxy for the overall chemical feedstock fraction) or derived from biomass, with the latter constrained by sustainable supply potentials.

The framework is further extended to evaluate alternative transition mechanisms, including intra-European relocation of industrial production, imports of green intermediate goods (HBI, ammonia, and methanol), and comparisons between domestic production costs and import prices. These extensions allow to quantify cost-competitiveness gaps and the subsidy levels required to retain industrial production in Europe under different decarbonization pathways. Due to space constraints, a detailed description of the industrial process representations, technology options, and techno-economic assumptions is provided in supplemental materials section S1, which documents the modeling choices underlying each industrial sector in full detail.

Additionally, this study integrates climate-adjusted weather projections into PyPSA-Eur. Standard PyPSA-Eur uses ERA5 (2013) and SARAH-3 data for wind and solar generation,[42] but climate change affects renewable generation, demand, and infrastructure. We use a dataset from Antonini et al.,[43] which combines historical ERA5 data (1940–2023) with CMIP5 EURO-CORDEX projections (2006–2100) under Representative Concentration Pathways (RCP) 2.6, 4.5, and 8.5. It provides country-level time series for wind, solar, and hydropower across the EU27 (excluding Cyprus and Malta), UK, Norway, Switzerland, and Serbia, incorporating multiple climate models to capture uncertainty.

## Parameters and assumptions

### *Common assumptions*

A consistent set of parameters is applied across all scenarios. PyPSA-Eur is operated in myopic mode with a 3-hourly temporal resolution for the years 2030, 2040, and 2050 using Gurobi solver. The time resolution was chosen to balance computational tractability with temporal detail in a high-dimensional, sector-coupled optimization problem, and is consistent with recent large-scale European studies.[28,37] The model spans 39 nodes across 34 European countries, covering all EU27 member states (excluding Cyprus and Malta), United Kingdom, Switzerland, Norway, Albania, Bosnia and Herzegovina, Montenegro, North Macedonia, Serbia, and Kosovo. Each country is represented by at least one node. The modeling framework follows a brownfield approach, incorporating existing power and heating generation, transmission infrastructure, industrial hubs, enhanced in this study with data on existing plants added to the original PyPSA-Eur model. Renewable generation from onshore and offshore wind and solar PV can expand based on land eligibility via *atlite*, while hydropower is fixed and nuclear may increase if cost optimal. Electricity storage includes pumped hydro, batteries, and hydrogen systems (electrolyzers, tanks, and fuel cells). Power and methane grids use SciGRID gas[44] and OpenStreetMap.[45] Grid expansion is allowed, but line use is capped at 70% to approximate N-1 security. Due to delays in hydrogen pipeline deployment,[46] $H_2$ infrastructure is excluded, and each node meets demand via local production. A robustness analysis considers repurposing existing gas pipelines or investing in new ones (Figure S3). Most of the technology costs are updated for each optimization year using the technology-data package.[47] Weather-dependent resources are derived from year- and country-specific climate projections developed by Antonini et al.,[43] from CNRM-CERFACS-CM5 global model and downscaled with the CNRM-ALADIN63 regional model, based on the RCP 4.5 scenario. Sustainable biomass is capped at 1,372 TWh/a (JRC-ENSPRESO[48]), with no imports. $CO_2$ capture is limited to 50 Mt $CO_2$/a in 2030, 250 Mt $CO_2$/a in 2040, and 400 Mt $CO_2$/a in 2050, in line with EU targets and IAM projections.[49,50]

**Table 2. Scenario matrix combining climate targets and industrial production trends, relocation within Europe and intermediate imports**

| | Climate target | | Relocation | Intermediate imports |
|---|---|---|---|---|
| Industrial production | *no climate policy*; *continued decline*; no relocation; no interm. imports | ***c**limate policy*; *continued decline*; no relocation; no interm. imports | climate policy; continued decline; *relocation within Europe*; no interm. imports | climate policy; continued decline; no relocation; *intermediate imports* |
| | *no climate policy; stabilization*; no relocation; no interm. imports | *climate policy; stabilization*; no relocation; no interm. imports | N/A | N/A |
| | *no climate policy; reindustrial*; no relocation; no interm. imports | *climate policy; reindustrial*; no relocation; no interm. imports | Climate Policy; reindustrial; *relocation within Europe*; no interm. imports | climate policy; reindustrial; no relocation; *intermediate imports* |

### *Scenarios framework*

Different scenarios are illustrated in Table 2. In *no climate policy* scenarios, the system is cost-optimized without GHG targets, though renewable investments may occur if economically advantageous. *Climate policy* scenarios implement EU targets: 55% GHG reduction by 2030, 90% by 2040, and net zero by 2050,[1,30,51] with all GHG expressed in $CO_2$ equivalents. Sectoral assumptions differ: land transport is fully electrified, methanol meets 50% of shipping demand (boosting industrial methanol production), and aviation uses synthetic kerosene via Fischer-Tropsch synthesis. As a note, Fischer-Tropsch processes output a range of hydrocarbon products; in this study, naphtha is used as a proxy for chemical feedstocks, and process parameters are defined at the level of the overall product slate.

The second scenario differentiation concerns industrial production. European industry has declined due to high energy costs, aging infrastructure, and global competition,[52,53] prompting policies like RePowerEU,[54] the Clean Industrial Deal,[3] and the Carbon Border Adjustment Mechanism (CBAM)[55] to support competitiveness and the green transition. To capture uncertainty, three trajectories are considered: *continued decline* (continued decline), *reindustrialization* (annual growth), and *stabilization* (stable output). These scenarios refer to domestic production; in *continued decline*, reduced output is assumed offset by imports, raising questions about the carbon intensity of foreign goods, while *reindustrialization* may reduce import reliance and create exports. Changes in domestic consumption are not explicitly modeled, and a full assessment of trade-related emissions is beyond this study's scope. These industrial production trajectories should be interpreted as stylized scenario assumptions designed to explore different structural pathways for European industry rather than forecasts derived from economic modeling.

To model the rate of change in industrial production, historical trends are extrapolated by replicating each sector's annual absolute change, based on linear interpolation between the earliest and latest available data (2023 for most sectors, 2022 for cement and plastics, and 2021 for methanol) as shown in Equation 1.

$$\Delta P = \frac{|P_{latest} - P_{first}|}{t_{latest} - t_{first}} \quad \text{(Equation 1)}$$

In Table 3 values for each sector are detailed. Figure 7 shows historical production (dots) and projected trajectories for each sector under the three production scenarios. Here, methanol values reflect industrial demand only, but under *climate policy*, the model has an additional shipping demand of 7 Mt in 2030, 16 Mt in 2040, and 23 Mt in 2050.

The analysis also explores two additional scenario dimensions under *climate policy* for both *continued decline* and *reindustrialization*. The first examines intra-European relocation of industrial production. In *no relocation* scenarios, production remains fixed, reflecting inertia due to infrastructure, labor, supply chains, and regional policies. In *relocation within Europe* scenarios, the model optimizes plant locations for cost-effectiveness, capturing a potential "renewables pull," where industries move to regions with abundant low-cost renewable energy.[23,29]

**Table 3. Historical annual production changes for each subsector**

| Subsector | Annual production change $\Delta P$ (Mt/year) | Source |
|---|---|---|
| Iron and steel | 2.6 | World Steel Association[56] |
| Cement | 1.9 | Cembureau[57] |
| Ammonia | 0.32 | Eurostat[58] |
| Methanol | 0.09 | Eurostat[59] |
| Plastics | 0.63 | Plastics Europe[60] |

The second dimension examines imports of intermediate green goods under *climate policy* for both *continued decline* and *reindustrialization*. This strategy retains high-value industrial segments in Europe while outsourcing energy-intensive stages to regions with lower renewable costs.[23,28,61] Imports considered include green methanol and ammonia for 2040 and 2050, using literature-based average prices,[25,28] and green HBI for EAF at 395 EUR/ton.[24] These feedstocks also serve as low-carbon fuels in other sectors. The assumption that imports of green intermediate goods become available only from 2040 reflects the expected timeline for the large-scale deployment and export capability of low-carbon industrial production outside Europe.

### *Assumptions for imports of green intermediates*

Prices for imported green intermediates are based on Neumann et al.,[28] which derive levelized costs of imported goods using the TRACE model.[62] In this framework, complete supply chains are optimized in exporting regions, including renewable electricity generation, conversion technologies (e.g., electrolysis and synthesis), storage, and shipping, under consistent techno-economic assumptions. Methanol is assumed to rely exclusively on DAC as the carbon source, while ammonia imports do not require carbon inputs. The resulting import prices are implemented as exogenous supply options in PyPSA-Eur and are therefore directly comparable with domestically produced goods, which are modeled using the same techno-economic assumptions for the target year (2040 and 2050). Methanol and ammonia are considered to be transported as liquid fuels and HBI is solid, thus all green intermediate imports in this study are assumed to be freely distributed within Europe.

### *Government subsidies calculations*

Subsidies are modeled in this study as an analytical device to quantify the competitiveness gap between domestic low-carbon production and imported green alternatives, rather than as normative policy recommendations. The resulting subsidy levels should therefore be interpreted as indicative magnitudes of the cost differential that would need to be bridged to retain industrial activity within Europe under different transition pathways. Importantly, the analysis does not assume that such subsidies would be fully implementable under existing EU competition law or state aid rules. Instead, the subsidy estimates serve to highlight the scale of the challenge faced by industrial policy and to inform discussions on feasible combinations of trade, regulation, and targeted support instruments.

Subsidy requirements are quantified, for each commodity $c$ in scenario $s$ and year $y$, by multiplying its production volume $Q_{c,s,y}$ by the difference between PyPSA-Eur weighted average marginal prices and green import prices from the literature[25,28] (Equation 2).

$$S_{c,s,y} = \left(p^{PyPSA-Eur}_{c,s,y} - p^{Green\ Imports}_{c,s,y}\right) \cdot Q_{c,s,y} \qquad \text{(Equation 2)}$$

Green industrial products from outside Europe are assumed to be available from 2040 (also as in Figure 2); thus, subsidies are not applied in 2030. The average annual subsidy for commodity $c$ in scenario $s$ (2035–2055) is computed as the weighted mean over representative years, with each year representing a decade, yielding Equation 3.

$$\overline{S_{c,s}} = \frac{10}{20} \sum_{y \in [2040,2050]} S_{c,s,y} \qquad \text{(Equation 3)}$$

## RESOURCE AVAILABILITY

### Lead contact

The lead contact is Alice Di Bella, who can be contacted by email at alice.dibella@cmcc.it.

### Materials availability

This study did not generate new unique reagents.

### Data and code availability

The code to reproduce the experiments is available at https://github.com/cerealice/pypsa-eur-adb/tree/industry; project and data are stored in a Zenodo repository: https://doi.org/10.5281/zenodo.20041935.

## ACKNOWLEDGMENTS

A.D.B. was funded by the European Union - NextGenerationEU, Mission 4, Component 2, in the framework of the GRINS - Growing Resilient, INclusive and Sustainable project (GRINS PE00000018 - CUP C83C22000890001). She has also received funding from the European Union's Horizon programme under grant agreement no. 101081604 (PRISMA). T.S. gratefully acknowledges funding from the Kopernikus-Ariadne project by the German Federal Ministry of Research, Technology and Space (Bundesministerium für Forschung, Technologie und Raumfahrt, BMFTR), grant no. 03SFK5R0-2. The views and opinions expressed are solely those of the authors and do not necessarily reflect those of the European Union, nor can the European Union be held responsible for them.

## AUTHOR CONTRIBUTIONS

A.D.B.: conceptualization, data curation, formal analysis, investigation, methodology, software, visualization, writing – original draft, and writing – review & editing. T.S.: data curation, formal analysis, software, and writing – review & editing. T.B. and M.T.: writing – review & editing, conceptualization, and project administration.

## DECLARATION OF INTERESTS

The authors declare no competing interests.

## DECLARATION OF GENERATIVE AI AND AI-ASSISTED TECHNOLOGIES IN THE WRITING PROCESS

During the preparation of this work the authors used ChatGPT in order to edit and improve the text of the manuscript. After using this service, the authors reviewed and edited the content as needed and take full responsibility for the content of the publication.

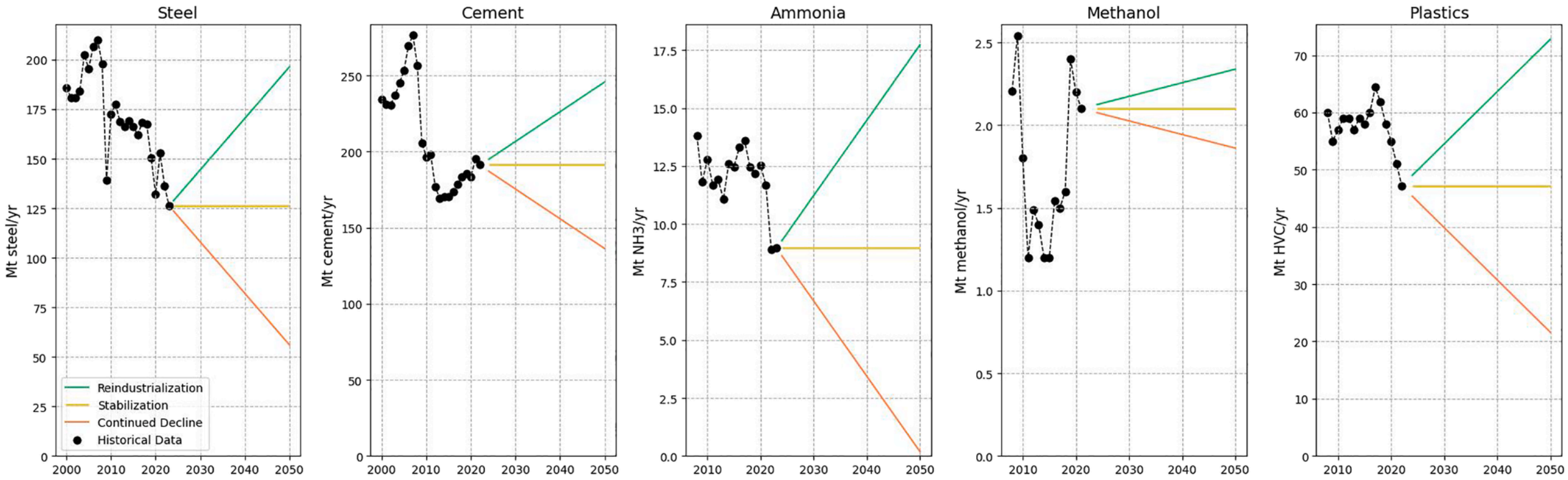

**Figure 7. Industrial production trajectory within the geographical scope of the model, EU27 (without Cyprus and Malta), UK, Switzerland, Norway, Albania, Bosnia and Herzegovina, Montenegro, North Macedonia, Serbia, and Kosovo, for the five industrial goods considered in the study, in the *continued decline*, *stabilization*, and *reindustrialization* scenarios**

Note that methanol production data exhibit an irregular trend, which may be attributable to inconsistencies or gaps in reporting rather than to underlying structural changes in the sector.

## SUPPLEMENTAL INFORMATION

Supplemental information can be found online at https://doi.org/10.1016/j.crsus.2026.100837.